# Theoretical Study of Elastic Properties of SiC nanowires of Different Shapes

Pavel B. Sorokin[1,2,3,4], Dmitry G. Kvashnin[1], Alexander G. Kvashnin[1], Pavel V. Avramov[1,3] and Leonid A. Chernozatonskii[2]

[1] Siberian Federal University, 79 Svobodny av., Krasnoyarsk, 660041 Russian Federation

[2] Emanuel Institute of Biochemical Physics, Russian Academy of Sciences, 4 Kosigina st., Moscow, 119334, Russian Federation

[3] Kirensky Institute of Physics, Russian Academy of Sciences, Akademgorodok, Krasnoyarsk, 660036 Russian Federation

E-mail: PSorokin@iph.krasn.ru

*The atomic structure and elastic properties of silicon carbide nanowires of different shapes and effective sizes were studied using density functional theory and classical molecular dynamics. The surface relaxation led to surface reconstruction with splitting of the wire geometry to hexagonal (surface) and cubic (bulk) phases. Theoretical calculations of effective Young's modulus and strain energies allowed us to explain the key experimental data of the SiC nanowires of different types.*

---

[4] Author to whom any correspondence should be addressed.

# 1. Introduction

Silicon carbide (SiC) is a very promising material to design effective high power, high frequency and high temperature semiconducting devices [1]. The high breakdown field, high thermal conductivity and high saturation velocity [2] concur in making SiC stand out from other wide band-gap materials (e.g., GaN, GaP, diamond).

Silicon carbide 1D nanoclusters are promising structures to realize new nanodevices with unique properties. The SiC nanotubes [3, 4] are an example of 1D SiC nanoclusters with tense atomic structure due to uncharacteristic SiC $sp^2$ hybridization. The 1D SiC nanowires (SiCNWs) overcome this disadvantage due to bulk crystalline nature and can be promising structures for future nanodevices.

SiCNWs can be synthesized by a reaction of silicon with carbon nanotubes. Dai *et al.* [5] used multi-wall carbon nanotubes in a vapor-solid reaction, yielding SiCNWs and other carbide nanorods with effective diameters ranging from 2 to 30 nm. Zhang et al. [6] used single-wall nanotubes (SWNTs) in a solid-solid reaction with the control of the growth process to fabricate SWNT-SiCNW heterostructures with well-defined crystalline interface. SiCNWs of 10 to 30 nm were synthesized by chemical vapor deposition (CVD) [7, 8] by Zhou *et al.* [9] managed to reduce the nanowire size down to 5 nm using iron particles as the catalyst. Synthesis of SiCNWs with effective diameter less than 10 nm by arc-discharge [10, 11] or by direct chemical reaction [12] was also demonstrated. More recently, Yang et al. [13] proposed the synthesis of SiC nanorods by thermal decomposition of a polymeric precursor with thicknesses ranging from 80 to 200 nm. A promising achievement for nanowire-based molecular electronics is the growth of coaxial SiCNWs, reported by Shen *et al.* [14].

Recent experimental study of the mechanical properties of SiC nanowires reports the maximum bending strength of 53.4 GPa for SiC nanowire of 23 nm diameter [15], which is significantly larger than the comparative maximum value of 28.5 GPa, obtained for multiwall carbon nanotubes.

In the work of Menon *et al.* [16] branched clathrate nanowires were investigated. The stability and electronic properties were studied by semi-empirical quantum-chemical method.

To the best of our knowledge there are only few papers devoted to the theoretical study of the elastic properties of SiC nanowires. For example, the $\langle 110 \rangle$ and $\langle 111 \rangle$ SiCNWs were calculated in the papers by Makeev *et. al.* [17] and Wang *et. al.* [18], respectively. The buckling behavior of the nanowires was observed and Young modules were estimated.

Branched nanowires of various types offer another approach to increase structural complexity and physical properties [19]. In the paper by Wang *et al.* [20] a new way to synthesize hyperbranched Si and GaN nanowires was proposed. Alivisatos *et al.* [21] reported chemical wet synthesis of tetrapod (or branched nanocrystals) of cadmium telluride with the control of effective diameter of identical arms. Also, dendrite silicon wires [22] and branched SiC nanowires [23] were synthesized experimentally.

The main goal of this work is a theoretical description of elastic properties of $\langle 110 \rangle$ oriented silicon carbide nanowires having different shapes. The planewave pseudopotential DFT-LDA technique was used to calculate the elastic properties of a pristine SiC nanowire. The classical molecular dynamics was used to study the elastic properties of branched SiC nanowires. Theoretical description of 1D SiC structures allowed us to elucidate the mechanism of elastic and inelastic distortions under stress and calculate effective Young's modules of the structures. This study is organized as follows: Section II describes the computational methods and objects under investigation, followed by the results and discussion in Section III. Conclusions are presented in Section 4.

## 2. Structural models and methods of calculations

A branched 3C type Y-shaped (YSiCNW) nanowire was designed by connection of $\langle 110 \rangle$ oriented SiC stem with two $\langle 100 \rangle$ SiC branches with a typical crystallographic 90° angle between them (Fig. 1a). Several cluster models with different effective diameters and lengths of the stem and branches were designed.

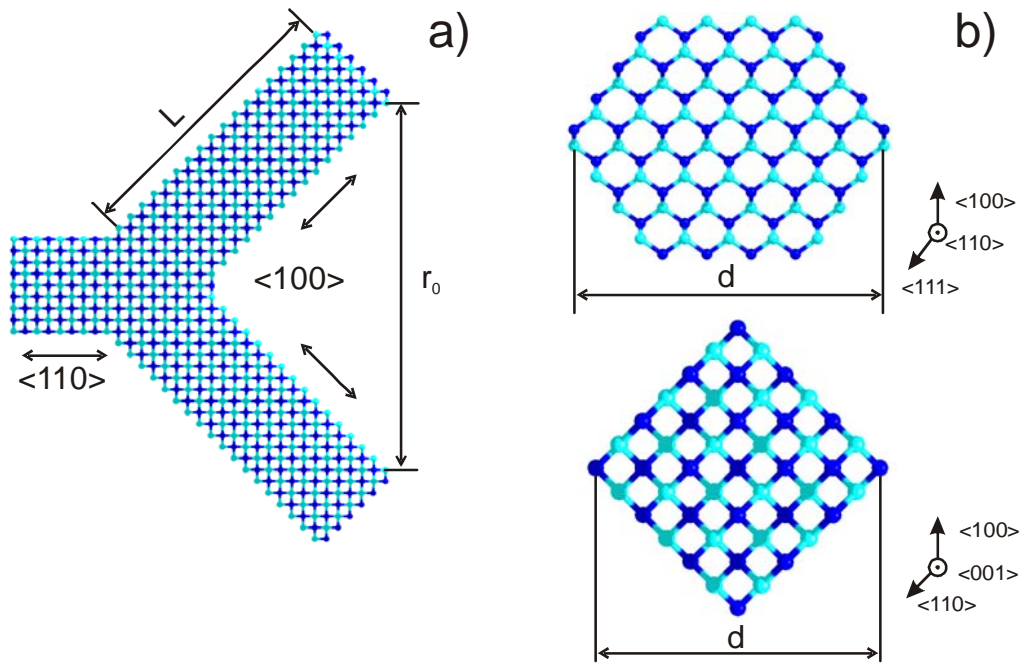

**Fig. 1. a) Branched YSiCNW. *L* is the branch length, *r*₀ is the initial distance between the branches b) the perpendicular cross-section in $\langle 110 \rangle$ and $\langle 001 \rangle$ wire directions studied in this work. Effective diameter, *d*, is estimated as the maximal distance between atoms on the opposite sides of wire cross-section.**

The electronic structure calculations of a set of silicon carbide nanowires were carried out using density functional theory in the framework of local density approximation [24, 25] with periodic boundary conditions using Vienna Ab-initio Simulation Program (VASP) [26-28]. We used a planewave basis set, ultrasoft Vanderbilt pseudopotentials [29] and a planewave energy cutoff equal to 358.4 eV. To calculate equilibrium atomic structures, the Brillouin zone was sampled according to the Monkhorst–Pack [30] scheme with a 1×1×8 k-point convergence grid. To avoid interactions between the species, neighboring SiCNWs were separated by 10 Å in the tetragonal supercells. During the atomic structure minimization, structural relaxation was performed until the change in total energy was less than $10^{-4}$ eV/atom. The calculated elastic module $C_{11}$ of bulk 3C-SiC (380 GPa) is in close agreement with the previous experimental (390 GPa [31], 363 GPa [32]) and theoretical (371 GPa [32]) results. The atomic geometry was predicted with an error of 0.2% in the crystal lattice constant ($a_{exp}$ = 4.3596 Å [33] and $a_{theory}$ = 4.37 Å). To study the effect of wire surface reconstruction (the formation of dimers on the wire surface between atoms in the neighboring cells) we calculate the double SiCNW cell.

To calculate the atomic structure and elastic properties of SiC branched nanowires classical molecular dynamics with Tersoff potential [34] was used. The potential is known to describe adequately the crystalline [35], amorphous [36] and nanostructured [37] phases of SiC and its nonstoichiometric alloys [38].

The method of atomic plane [39, 40] was used to simulate external pressure. Previously this method was used to study the elastic properties of multiterminal carbon nanotubes [39, 40]. The bending strain on the junctions of the YSiCNWs was created by a piece of atomic plane placed parallel to the stem axis. We choose the pure repulsive potential between the plane and nanowire for avoiding nonrealistic bonding between them. The plane was driven towards the branch in small steps and the YSiCNW was optimized in each step (see Fig. 2). The end of the YSiCNW stem was fixed.

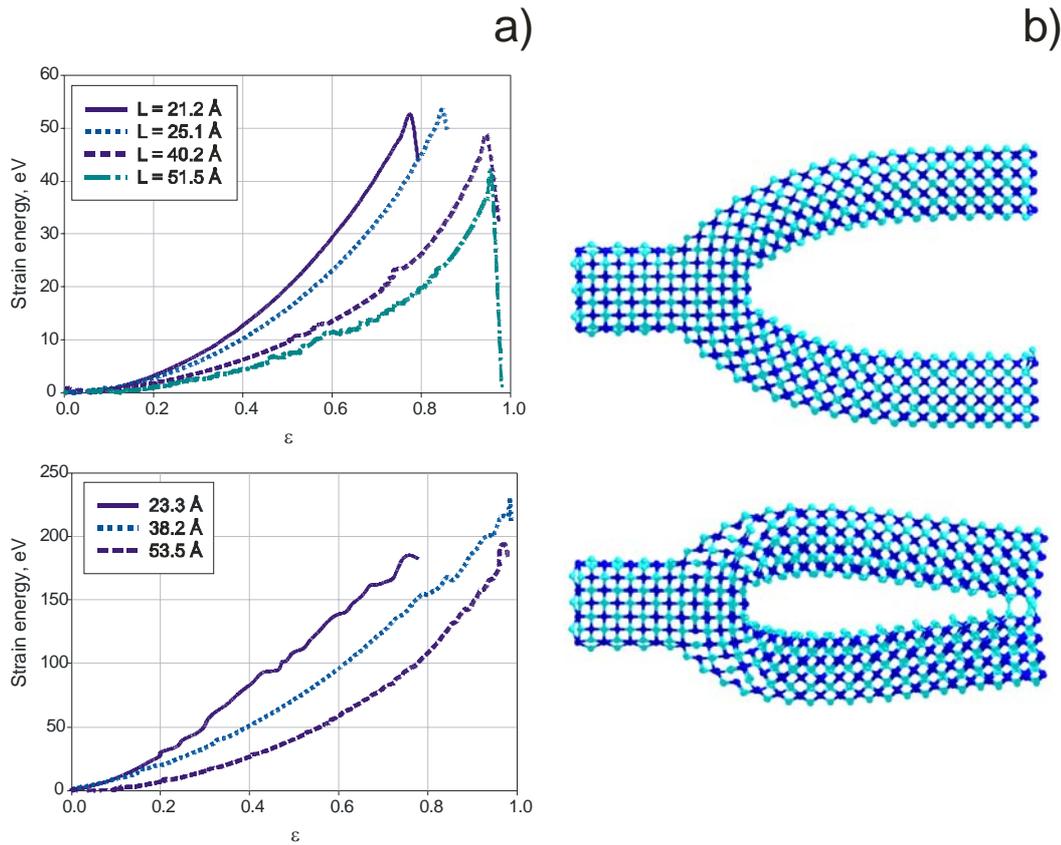

**Fig. 2. a) The strain energy vs strain plotted for the YSiCNW with effective diameter 11.9 Å (top) and 17.2 Å (bottom) with various branch lengths shown in the figure legend; b) the strained YSiCNW structure ($L$ = 40.2 Å, $d$ = 11.9 Å) in elastic regime ($\varepsilon$ = 0.69, top) and in beyond the critical point ($\varepsilon$ = 0.97, bottom). The formation of new bonds between branches decreases the energy and changes the atomic geometry.**

We defined the bending strain as $\varepsilon = \frac{\Delta r}{r_0}$ where $\Delta r = r - r_0$ ($r_0$ is the distance between branch ends in the relaxed structure and $r$ is the distance in the strained one, Fig. 2a). The effective Young module of the branched wires was estimated as $Y_{eff} = \frac{E''}{r_0 S}$, where the strain energy $E$ is approximated as $E = \frac{1}{2} E'' \varepsilon^2$ on the assumption that the loading is created by the atoms of the atomic plane ($S = S_{fragment}$ is the square of the plane equal to 39.3×19.9 Å$^2$).

The strain is applied along $z$ direction in increments of +1 % (tension) and −1 % (compression). The strain energy is the total elastic energy accumulated in the nanowire due to external loading. In the elastic regime, the elastic energy depends on the applied strain ($e_{zz}$) quadratically and proportional to one half of the Young's module. Consequently, the Young's module can readily be extracted from the theoretical data using $Y = \frac{1}{V_{atom}} \frac{\partial^2 E_{strain}}{\partial e_{zz}^2}$ equation, where $Y$ is Young's modulus, $E_{strain}$ is the strain energy per atom and $V_{atom}$ is the normalized system volume [17], $V_{atom} = \frac{V_{bulk}}{N_{atoms}}$. $V_{bulk}$ is the volume of the SiC-3C unit cell, $N_{atoms}$ is the number of atoms in the cell. The definition of the $V_{atom}$ gives us a possibility to make a qualitative assessment of the nanowire volume.

**3. Results and discussion**

The perpendicular cross-section of initial and relaxed structures of SiCNW calculated using DFT-LDA method is presented in Figs 2a and 2b. The [100] surface reconstruction is caused by the formation of Si-Si and C-C dimers on the [100] facets (Fig. 3b) as well as due to increasing the distance between the first and the second layers of the SiCNW (Fig. 3a). Similar effect was studied in the paper by Ivanovskaya *et al.* [41] for BN nanowires and Barnard *et al.* [42] for diamond NW. The nanowire structure consists of the outer hexagonal layer and internal crystalline rod with the cubic 3C-phase structure of SiC. Actually, the SiC nanowire atomic structure consists of $sp^2 + sp^3$ phase.

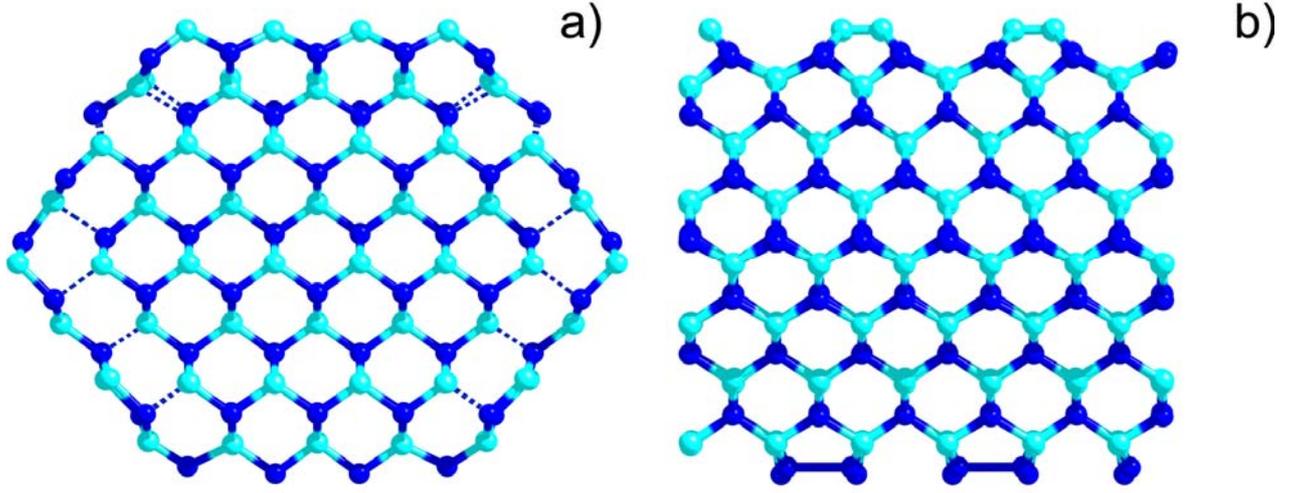

**Fig. 3. a) Cross-section and b) side view of relaxed $\langle 110 \rangle$ SiCNW with effective diameter d = 21.64 Å. Chemical bonds longer 1.95 Å are marked by the dashed lines.**

In the paper by Sun *et al.* [43] the SiC multi-wall nanotubes were observed. Menon *et. al.* [3] predicted the hexagonal SiC nanotubes and calculated the single hexagonal SiC layer. The distances between silicon and carbon in the layer are close to the distances between surface atoms of the studied wires (Table 1). The unit cell parameter, *c*, of the wires tend to the bulk 3C-SiC value (which is equal to the length of the diagonal of base of the cell $\sqrt{2}a$).

**Table 1. The atomic structure of the $\langle 110 \rangle$ SiCNW**

| *d* (Å) | Si-C distance of the surface atoms (Å) | Si-C distance between surface atoms and outer atoms in the rod (Å) | SiC distance in the rod (Å) | *c* (Å) |
|---|---|---|---|---|
| 9.19 | 1.74-1.92 | 1.90-2.39 | 1.84-1.91 | 6.18 |
| 15.56 | 1.74-2.04 | 1.89-2.16 | 1.84-1.97 | 6.175 |
| 21.64 | 1.74-2.75 | 1.84-2.36 | 1.85-1.97 | 6.168 |
| Bulk | 1.78-1.81 (hexagonal phase) [3] | - | 1.89 (cubic phase) [33] | 6.16 ($\sqrt{2}a$) [33] |

Comparison of our DFT-LDA results with the theoretical [17] and experimental [44] data (

Table 2) shows that the Young's modules obtained in our calculations significantly differ from the early reported theoretical results [17]. We suppose that the differences are caused by different approaches used to calculate the wires (ab initio DFT in this paper and the classical Tersoff potential in the paper by Menon *et al.* [17]). The simulations of the wires with Tersoff potential do not lead to the appearance of the mixed $sp^2 + sp^3$ phase with increasing the wire stiffness.

Table 2. Elastic characteristics (Young's modulus) of the $\langle 110 \rangle$ SiCNW. The values are shown for nanowires with three different diameters *d*.

| $d$, Å | DFT-LDA (GPa) | Theory [17] (GPa) (diameter, Å) | Experiment [44] (GPa) (diameter, Å) |
|---|---|---|---|
| 9.19 | 440.0 | 504.8 (8.9) | |
| 15.56 | 442.4 | 525.5 (14.2) | 660 (175) |
| 21.64 | 445.8 | 537.4 (19.6) | |

The changes in the total energy in comparison with the total energy of the initial strain-free configuration reflect the strain energy $E$ as a function of strain $\varepsilon$. At small strain the strain energy of the structures displays quadratic behavior $E = \frac{1}{2} E'' \varepsilon^2$ (Fig. 2a).

Increasing the pressure leads to increasing the bond lengths between neighboring atoms in the outer wire surface region and shortening of the ones in the inner surface region with a visible deflection of the bond angles from their natural tetrahedral value of 109.471º (compare Fig. 2b top and bottom). The region of the main distortion of the crystalline structure is the branch crossing interface. Further increasing the tension leads to reforming the chemical bonds in the region of branch crossing and the formation of new bonds between the branch ends. The lattice structure of other nanowire regions remains practically undistorted.

We estimated the effective Young module of the YSiCNW junctions with effective diameters 11.9 Å and 17.2 Å (Fig. 4). The elongation of the branches from 21.2 to 70.0 Å and from 23.3 to 71.4 Å leads to decreasing the Young module from 1.46 GPa to 0.07 GPa and from 4.85 GPa to 0.59 GPa, respectively. Due to decreasing the relative number of distorted bonds and bond angles with increasing branch lengths, the effective Young module of the wires with longer branches can not be larger for the studied systems.

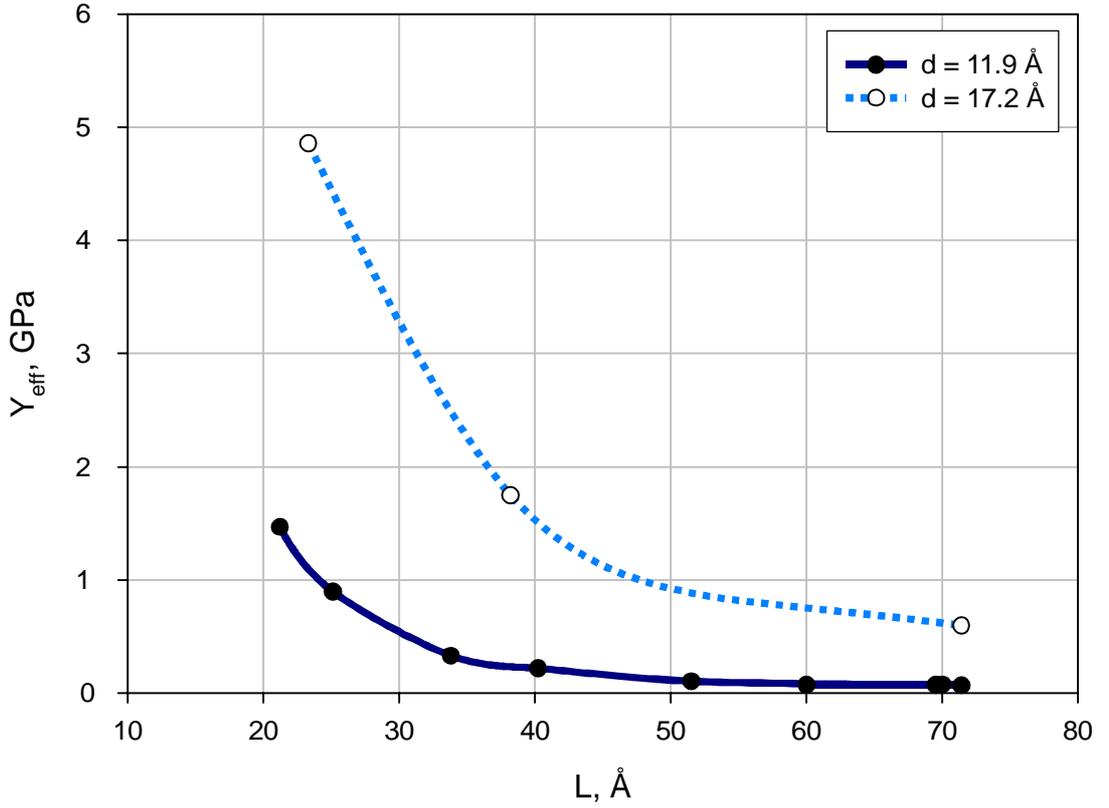

**Fig. 4. . The effective Young's modulus ($Y_{eff}$) for YSiCNW as a function of branch length ($L$) of the nanowires. Two curves correspond to different nanowire diameters (11.9 and 17.2 Å).**

Let us compare the stiffness of the branched SiC nanowires with muliterminal nanotubes [40]. For Y-type single wall nanotubes ($L = 45$ Å, branch effective diameter $d = 12.2$ Å) and YSiCNW ($L = 51.5$ Å, $d = 11.9$ Å) the $E''/r_0$ values are equal to 0.24 eV/Å and 0.59 eV/Å, respectively. The $E''/r_0$ value was used because it was independent from the square of the atomic plane which is different from the one in Ref. 40. These results are obvious because nanowires have the solid crystal structure whereas nanotubes are hollow nanomaterial.

The passivation of the surface of the wire changes the mechanical behavior especially for the critical values of bending. The passivating atoms (i.e. hydrogens) prevent the bonding between the branches and therefore increase the flexibility of the wire. This effect was observed in the case of silicon branched wires [45]. Also the passivated wires should have bigger effective Young module due to the interaction of neighboring passivating atoms during the bending.

The unloading of elastically bent wires should lead to oscillations of branches with the frequency depending upon the branch lengths and diameters and the structures can be possibly used as tuning forks for ultrahigh frequencies.

## 4. Conclusions

The elastic properties of silicon carbide nanowires of different shapes were studied by using density functional theory and molecular dynamics simulations with Tersoff model potential. It was found that the surface relaxation led to significant surface reconstruction with splitting of the wire geometry to hexagonal (surface) and cubic (bulk) phases. The Young modules of the wires were calculated.

The bending process of branched silicon carbide nanowires was investigated. In inelastic regime the formation of new bonds between different parts of the nanowires in the region of branch crossing was observed. The dependence of the effective YSiCNW Young modules on the branch lengths was found. The stiffness of the wires and nanotubes from the reference paper was compared. It was found that the stiffness of the wires was much larger due to crystalline structure of the wires.

**Acknowledgements**

This work was partially supported by CREST project and by the Russian Foundation for Basic Research (grant n. 09-02-00324 and 09-02-92107). The electronic structure calculations have been performed on the Joint Supercomputer Center of the Russian Academy of Sciences. One of the authors (PVA) acknowledges the encouragement of Dr. K. Morokuma, research leader of Fukui Institute. The geometry of all presented structures was visualized by ChemCraft software [46].